# VEHICLE SECURITY: RISK ASSESSMENT IN TRANSPORTATION

Kaveh Bakhsh Kelarestaghi, Mahsa Foruhandeh, Kevin Heaslip, Ryan Gerdes

*Abstract* - Intelligent Transportation Systems (ITS) are critical infrastructure that are not immune to both physical and cyber threats. Vehicles are cyber/physical systems which are a core component of ITS, can be either a target or a launching point for an attack on the ITS network. Unknown vehicle security vulnerabilities trigger a race among adversaries to exploit the weaknesses and security experts to mitigate the vulnerability. In this study, we identified opportunities for adversaries to take control of the in-vehicle network, which can compromise the safety, privacy, reliability, efficiency, and security of the transportation system. This study contributes in three ways to the literature of ITS security and resiliency. First, we aggregate individual risks that are associated with hacking the in-vehicle network to determine system-level risk. Second, we employ a risk-based model to conduct a qualitative vulnerability-oriented risk assessment. Third, we identify the consequences of hacking the in-vehicle network through a risk-based approach, using an impact-likelihood matrix. The qualitative assessment communicates risk outcomes for policy analysis. The outcome of this study would be of interest and usefulness to policymakers and engineers concerned with the potential vulnerabilities of the critical infrastructures.

## I. Introduction

IN the Wired article that broke the hacking of a Jeep Cherokee, researcher Charlie Miller was quoted as saying, "When you lose faith that a car will do what you tell it to do, it really changes your whole view of how the thing works." [1] The high demand for cutting-edge technologies and the need to keep vehicles affordable are reasons why rigorous security measures and policies have not been prioritized in the past. Vehicles are becoming more intelligent with the constant demand for improvement of driver's safety, comfort, and vehicle performance. Technologies are being deployed such as infotainment and safety monitoring applications in the vehicles to meet the demand for increased technology. According to the major automobile original equipment manufacturers (OEMs), nearly 100% of new vehicles are equipped with wireless technologies [2]. The security challenges become even more complex by the knowledge that 60% of the 2016 vehicles come with internet connection capabilities [3].

The vehicle's systems were designed without remote connectivity in mind and focused on vehicle safety and not security. Car manufacturers achieved the former but not in the presence of an attacker in an adversarial environment. Current vehicles are computers on wheels equipped with Wi-Fi access points, Bluetooth modules, and dozens of Electrical Control Units (ECUs). Even further, cars are envisioned as platforms to offer on-demand technologies and services to the road users. Advancements that could transform cars to have the functions of personal computers and smartphones, but with numerous security vulnerabilities [4].

The system should provide prevention strategies in addition to detection and recovery strategies to stop adversaries from causing any harm to ITS critical infrastructure. Such prevention strategies need to be built into the design from inception. Reasons to secure the in-vehicle network prior to implementation are [5]: (i) design phase is the most effective stage to prevent exposure, (ii) initial security consideration will prune research initiatives that are expensive to be secured once implemented, (iii) blind eye to attackers will endanger the system robustness, and (iv) security is crucial to government and customers. The real challenge emerges when an intelligent vehicle is placed into the vulnerable vehicular ad hoc network. An adversary can gain access to the in-vehicle network and be rewarded by ample access to the vehicle ECU. The adversary can take control of the vehicle braking system, engine, air conditioning system, and steering wheel are just steps needed to remotely-drive a vehicle.

There is a gap in the current literature to assess and evaluate the impact of cyber/physical attacks on the road users and the system operators. This study attempts to survey the impacts of in-vehicle network security vulnerabilities and adds to the literature of ITS security and resiliency, by exploring possible consequences of an adversary compromising the in-vehicle system. To this extent, we aim to contribute to the nascent but growing literature of ITS security impact-oriented risk assessment. First, most of the previous studies have been limited to combine risks of a compromised in-vehicle network systematically. The bulk of this work is devoted to identifying security attacks against the in-vehicle network and to proposing defense and protection security mechanisms. The knowledge about the conceivable attacks and the performance ᴷB. Kelarestaghi and K. Heaslip are with the Department of Civil & Environmental Engineering, Virginia Tech, Arlington, VA, 22203. Email: {*kavehbk, kheaslip*} @vt.edu.

M. Foruhandeh and R. Gerdes are with the Department of Electrical & Computer Engineering, Virginia Tech, 900 N Glebe Rd, Arlington, VA, 22203. Email: {*mfhd, rgerdes*} @vt.edu.

of the security mechanisms is still rudimentary. This deficiency urges the need for an impact-oriented analytic approach to identify system-level risk –that is an area of emerging research. A rigorous quantitative assessment could be conducted, but this is a difficult task because statistical data (collected through either revealed or stated method) is mainly unknown. As in-vehicle security impact assessment in transportation is active research with lacking numerical data of the executed security attacks, a comprehensive quantitative risk assessment methodology cannot be conducted. That is, in our second contribution we propose employing a qualitative risk-based approach for measuring risks and communicating the results for policy analysis. We employ National Institute of Standard and Technology (NIST) [6] risk-based model to conduct a qualitative vulnerability-oriented risk assessment. The risk-based approach of this study aims to identify adverse impacts that are caused by malicious adversaries exploiting in-vehicle network security vulnerabilities. In our third contribution, we provide a rating of the risks' likelihood and impact through a qualitative risk-based approach. Herein, we synthesize evidence from current literature and real-world hacking incidents to determine potential impacts of exploiting in-vehicle network vulnerabilities. We then map safety, operation, reliability, and security issues ensued by in-vehicle hacking incidents into a visual, matrix for risk prioritization purposes.

The remainder of this study is structured as follows. First, we explain the risk assessment methodology of this study. Second, we investigate the threat events and threat sources to the in-vehicle network. Third, we evaluate the in-vehicle network vulnerabilities and determine the risks of such vulnerabilities to the transportation system. We conclude our study by suggesting countermeasures to prevent the failure of ITS critical infrastructure.

## II. RISK ASSESSMENT METHODOLOGY

The risk assessment methodology of this study incorporates the risk model proposed by NIST Special Publication 800-30 (SP 800-30) [6]. The NIST risk assessment objective is to detect, evaluate and to prioritize risks to the system operations and assets [6]. In which the aim is to pinpoint insights to policymakers to support risk responses. To this extent, risk determination is conducted through identifying threats to the system, system vulnerabilities, consequences in case an adversary exploits those vulnerabilities, and the likelihood that impacts emerge. The deliverables of the risk assessment methodology will be the critical components for the risk management process [7]. The risk assessment is the second component of the risk methodology hierarchy that incorporates knowledge into the risk respond, the third component of the risk management (Figure 1). The four core elements (i.e., frame, assess, respond, and monitor) of the hierarchy constitute a comprehensive risk management process that ensures successful risk-based decisions that manage system's risks from strategic to tactical level.

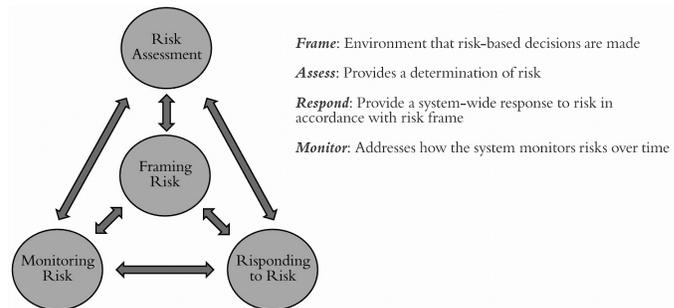

Fig. 1. Risk Management Process (adapted from [7])

For this study, we employ the risk model proposed in NIST SP 800-30 to conduct qualitative vulnerability-oriented threat analysis. The analytic approach of the risk assessment comprises of both qualitative assessment and impact-oriented analysis approach. Together, the qualitative assessment and impact-oriented analysis provide the level of detail to characterize adverse impacts for which likelihoods are identified. The risk-based approach of this study (Figure 2) includes (i) identifying threat sources and events, (ii) identifying system vulnerabilities and predisposing conditions, and (iii) determining adverse impacts and magnitude of impact.

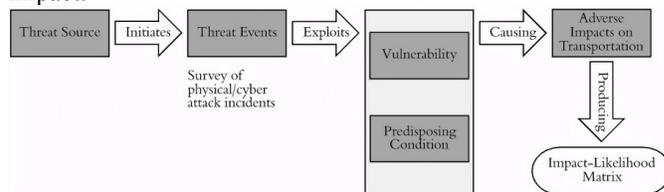

Fig. 2. Risk Model (adapted from [6])

## III. THREAT SOURCE AND EVENTS

"No matter what happens, don't panic," said Charlie Miller to Andy Greenberg before he hacked 2014 Jeep Cherokee through the car entertainment system [1]. The most analogous scenarios with real-world settings are Miller and Valsak experiments [1], [8]–[11], wherein a driver was behind the wheel during the attack. In 2013, they hacked Ford Escape and Toyota Prius through a physical attack. They demonstrated their work in a real-world situation when a driver was behind the wheel. Several attacks were successfully executed such as killing the engine, disabling the brake, and altering the speedometer. Later in 2015, Miller and Valsak remotely hacked the Jeep Cherokee with a driver driving the car [1]. This hack occurred by rewriting the firmware to gain access to the car Controller Area Network (CAN) bus. Examples of their attacks are (i) control of the engine, (ii) control of brake, (iii) control of steering wheel, and (iv) control of air conditioning system. The most important outcome of this attack is the reaction of the driver to the experience; the driver expressed his feeling as a "traumatic experience." The experience was traumatic because the driver did not have any control over the vehicle while driving on the highway surrounded by other vehicles. The driver was helpless when hackers turned radio volume to the maximum volume and left him with no choice but to ask them to make it stop [1].

Before the Miller and Valsak studies [9]–[11], other researchers have exploited vulnerabilities of the vehicle systems that are based on the CAN bus technology. Hoppe et al. [12], indicated that by exploiting CAN bus security vulnerabilities, an adversary is capable of (i) compromising the availability of window lifts through DoS attack, (ii) compromising the availability of the warning light by keeping it dark, and (iii) removing airbag control modules from the system. They argued that an adversary could cause passengers discomfort and compromise their safety. Koscher et al. [13] examined the security of two 2009 model year passenger cars (researchers did not disclose the make and model of the cars) through reverse engineering, in both lab and test road settings. The researchers demonstrated that by taking control of even one of the car's ECU an adversary can entirely compromise the car and that ECU can be reprogrammed even while driving. They were able to take control of several car components by hacking the vehicle (i) physically through the On-Board Diagnostic II (OBD-II) port, and (ii) remotely (with prior physical access) through car wireless interfaces (digital radio). These components include (i) radio, (ii) instrument panel cluster, (iii) body controller (e.g., door lock, trunk lock, and window relays), (iv) engine, (v) brakes, and (vi) heating and air conditioning. Later in 2011, Checkoway et al. [14] supplemented the Koscher et al. [13] experiment by attacking a modern vehicle remotely, without prior physical access. They were able to compromise the vehicle through several components entirely. These components include (i) diagnostic tool, (ii) media player, (iii) Bluetooth, and (iv) cellular communication. The dreadful reward of an adversary breaching through the wireless components is that he does not need to execute an attack immediately and can wait for a time to gain maximum benefit.

Other researchers have also attempted to scrutinize the security of vehicles. Oka et al. [15] suggest that adversaries can breach into the in-vehicle network through the Bluetooth communication. They reviewed publicly available information and concluded that although it is difficult to pair with the built-in infotainment system, it is possible to compromise driver safety through the external infotainment systems via Bluetooth. It is even more likely for an attacker to inject false messages through the OBD-II port dongle by merely pairing his device or with a brute-force attack. Another study conducted by Foster et al. [16] assessed the security vulnerabilities of the aftermarket Telematics Control Units (TCU) that interconnect ECUs with external systems. They examined two threat models: (i) local (i.e., physical access to the TCU) and (ii) remote in which the attacker can compromise the vehicle without physical access. The results indicate that in both cases adversaries with adequate knowledge can compromise the vehicle through the TCU. The adversary can inject false CAN packets and fully control safety components of a vehicle such as a brake system. The researchers tested this attack in a real-world situation and were able to tamper with windshield wipers and the brake system.

The Yan et al. [17] study indicates that it is conceivable to hack Tesla Model S sensors through the spoofing and jamming attacks. The study suggests that these attacks can trigger a crash and compromise the function of the self-driving vehicles. The study has some limitations such as the limited range of the attacks and the uncertainties that come with the possibility of on-road attack. Nevertheless, the impact of such attacks on road users' safety is considerable and needs to be further assessed. These vulnerabilities are not the concern only of passenger cars but involve heavy commercial vehicles as well. Burakova et al. [18] assessed the security of a 2006 Class-8 semi-tractor and a 2001 school bus that are both equipped with SAE J1939 standard; the J1939 bus is commonly used among the heavy-duty vehicles in the US, for in-vehicle communication and diagnostics. The study results indicate that an attacker with network access can compromise the ECUs and manipulate (i) gauges on the instrument cluster (e.g., oil pressure and battery voltage gauge), (ii) engine Revolutions Per Minute (RPM), and (iii) engine brake. In conclusion, the attacker with physical access to the in-vehicle network may override the driver and remotely control the heavy-duty vehicle.

Jafarnejad et al. [19] were able to hack the Renault Twizy 80 through exploiting vulnerabilities of the Open Vehicle Monitoring System (OVMS); the OVMS enables a driver to have remote access to electric car information such as battery level and location. They concluded that an adversary might override the driver and launch several attacks. These attacks are: (i) move the vehicle forward, and backward, (ii) change the speed of vehicle while in operation, (iii) cause damage to the engine, (iv) change motor direction at will, (v) distort a dashboard, and (vi) control the throttle. In 2015, white hackers physically breached into the Virginia State Police (VSP) Chevrolet Impala. A police officer who intended to start his duty was not able to shift the gear. The hacker was able to manipulate the speedometer and eventually shut down the engine. Tampering was initiated with the installation of a device in the car, capable of Bluetooth communication [20]. The experiment indicates that even a vehicle without design-built advanced communication is a lucrative target for adversaries. In a real-world situation, consequences of a malicious behavior targeting state trooper vehicles can cause delay and disruption of a police operation. This event provoked VSP to support security-based research and to simulate such an event on its students to prepare them for unexpected attacks. Also, a private company designed a real-time intrusion detection device to be plugged into the OBD-II [20]. Table 1 represents a summary of the experiments that attempted to hack on-the-market vehicles.

The review of the existing literature suggests that a malicious adversary should have high-level of expertise, well resourced, and be capable of executing effective, continuous, and synchronized cyber/physical attacks. Vehicle security vulnerabilities are not only of today or past concern; these can accrue with vehicular ad hoc network security issues and could trigger dire consequences. The breach might prolong and contaminate other vehicles that are on the same network as the hacked vehicle. Breach propagation can disseminate over the ITS wide-range network like a virus goes from one computer to another.

TABLE 1
SUMMARY OF THE IN-VEHICLE NETWORK EVENTS

| 1st author | Year | Vehicle | Foothold | Threat Model | Attack Impact |
|---|---|---|---|---|---|
| Burakova [18] | 2016 | 2006 Class-8 semi-tractor and 2001 school bus | OBD-II | Physical | Taking control of gauges on the instrument cluster, engine rpm, and engine brake |
| Yan [17] | 2016 | Tesla Model S | Sensors | Physical | Trigger a crash, Compromise self-driving functions |
| Higgins [20] | 2015 | Chevrolet Impala | OBD-II | Physical | The driver was not able to shift the gear, tampering the speedometer, and kill the engine |
| Greenberg [8], and Miller [10] | 2013 | Ford Escape, Toyota Prius | OBD-II | Physical | Disable brake, changed seatbelt setting, take control of steering wheel, spoofed GPS, distort a dashboard, and manipulate vehicle lights |
| Koscher [13] | 2010 | 2009 model year passenger car | OBD-II | Physical & Remote (with prior physical access) | Take control of radio; instrument panel cluster; body controller; engine; brakes; heating and air conditioning |
| Greenberg [1], and Miller [11] | 2015 | 2014 Jeep, Cherokee | Entertainment system –Cellular (Wi-Fi hotspot) | Remote | Take control of the engine, take control of brake. Take control of the steering wheel, control air conditioning system, traumatic driving experience, compromised driver privacy by tracing the location of the car through the GPS coordinates. |
| Jafarnejad [19] | 2015 | Renault Twizy 80 | OVMS | Remote | Move the vehicle forward and backward, change the speed of the vehicle, cause damage to the engine, changing motor direction, distort a dashboard, control the throttle |
| Checkoway [14] | 2011 | 2011 model moderately priced sedan | OBD-II, Media player, Bluetooth, and Cellular | Remote | Complete control over the car systems, car theft, and surveillance |

## IV. SYSTEM VULNERABILITY

Vehicles or "computers on wheels" are prone to cyber/physical attacks as a result of vehicle security vulnerabilities. A premium vehicle may consist of more than 70 ECUs that are interconnected through the network data bus. This network consists of a core bus, the CAN, and sub-networks which are: (i) Local Interconnect Network (LIN), (ii) FlexRay, and (iii) the Media Oriented System Transport [21]. CAN has been the de facto standard [22], [23], yet CAN vulnerabilities such as weak access control, no integrity and no authenticity, make it a potential foothold for adversaries to launch eavesdropping and replying attacks aiming for ECUs [24].

One may claim that ECUs are hackable with significant effort through reverse engineering and carjacking, similar to the Koscher et al. study [13]. In fact, researchers have been demonstrating that attacks are possible with more ease remotely via Bluetooth, Cellular radio [14], [24], or through a malicious smartphone application [24]. In these attacks, although the physical access does not need to be granted, an attacker needs high-level knowledge of vehicle's network architecture to hack the vehicle.

A valid concern is that a sophisticated adversary can breach into the vehicular ad-hoc communication network and inject false information to an in-vehicle network and be rewarded with access to the ECUs [25]. The consequences might be to disable a braking system, shut down an engine, distort a dashboard, and manipulate the air conditioning system, lights, radio, driver seat, to name a few [13], [14], [24]. Beyond all, the worst-case scenario might be that an adversary fully controls a vehicle or in other words, remote-drives the vehicle.

## V. ADVERSE IMPACTS

Researchers have paid a great deal of attention to the security aspect of vehicular network protocol and structure, but unfortunately, very little attention has been dedicated to assessing the impact of vehicle security breach on the transportation network. Although many studies suggest the presence of various security vulnerabilities within the in-vehicle network, surprisingly, cyber/physical attacks impact assessment has not been extensively explored. Thus, we cannot say for sure what consequences such an attack can impose. This section represents possible adverse impacts of in-vehicle network security breach. The implications are based on real-world studies that have focused on or alluded to the vehicle cyber/physical attack impacts on transportation.

Characteristics of the intelligent vehicles are a "double-edged sword." Intelligent advancements offer substantial applications, but on the other hand, it stimulates security threats and malpractices. With the developments in the age of the ubiquitous Internet and cellular communication soon the conventional vehicles will be part of the Internet of Things (IoT), a new era that is called the Internet of Vehicle (IoV). It has been projected that approximately between 19 billion to 40 billion of "things" will be connected through the Internet, and vehicles will constitute a substantial portion of the IoT [26]. With the emergence of the IoV and the progression of technology, adversaries will not limit themselves to the known attacks, but they will devise innovative attacks to exploit IoV vulnerabilities. The adaptiveness of the adversary promotes the need for more comprehensive strategies and out of the box perspective.

While manufacturers, legal authorities, and service providers are the core pieces of the transportation network, the ones who are disturbed the most by the security breaches are the road users. Examples of this scenario are the recent Yahoo [27] and Equifax [28] breaches, where privacy and security of millions of people were undermined, due to the system footholds. Attacks such as location tracking can jeopardize users' privacy and impose problems for the road users. Also, the attack on vehicle security can lead to a sophisticated leak of private information. That is, by compromising the infotainment system, an adversary could eavesdrop passengers' communications through the in-cabin microphone, and extract that data via the connected Internet Relay Chat (IRC) channel [14]. Even minor information on specifications of the sensors, used in the physical infrastructure of a vehicle, impart reliable information on the make and model of the vehicle [29]. In the Vehicular Ad hoc Network (VANET), vehicles are configured to send timestamps, identifiers and hello beacons periodically to communicate with other vehicles. Eavesdropping this information could trigger sophisticated privacy concerns than location tracking [29]. For example, through reading the hello beacons associated with a specific identifier, an adversary could gain access to information related to individuals' arrival and departure. Unauthorized investigation parties can also use this information that can impose severe consequences.

Imperative to note that there is no clear way of defining privacy and setting its boundaries. The degree of privacy is dependent on many factors such as user preferences, environmental settings, and application. Sometimes privacy can even be entangled with regulations of a country. This ambiguity is the main reason for the lack of considerable efforts to accommodate privacy in transportation systems, on a global scale. For instance, to fulfill certain levels of privacy, any necessary connections to some fixed physical or software infrastructure has to be avoided, which happens to be a hard task to be undertaken by the manufacturers at a technical level. Other than that, a substantial problem emerges when security and privacy happen to contradict each other in specific scenarios. The information breach on specific sensors can be an excellent example of this scenario. That is, sensor information might be used for designing intrusion detection mechanisms, while at the same time sensors can be a foothold for adversaries to compromise the system's privacy. Some frame the discussion of privacy between two extreme edges of full anonymity or no privacy at all. How to find the proper trade-off is still an open problem to be solved. A universal business model needs to be developed via an international effort of technical parties as well as legal authorities.

Disruption of the ITS network usually comes with monetary loss. These losses can affect both operators and users of the system. A sophisticated adversary can cause traffic congestion, driver confusion, driver distraction, and may be able to disrupt the system availability, which dictates unexpected operation cost [30] (e.g., time and labor cost). Schoettle and Sivak [31] surveyed 1,533 individuals and found that almost 70% of US participants are moderately or highly concerned about security and data privacy breaches when it comes to the automated vehicle technology. In fact, the presence of any vulnerabilities may lead to the point that public loses trust in the system.

As far as the consequences of in-vehicle network security vulnerabilities are concerned, several impacts can be singled out. These consequences are mainly concerned with transportation's safety, operation, efficiency, security, and reliability, and drivers' behavior. The hacking in-vehicle network may impose ruinous damages by compromising road users' safety. Safety concerns have been identified in several studies (e.g. [12]). Taking control of the engine, brake, steering wheel and throttle [10], [11], [19] to supersede the driver, compromises the safety of the targeted driver and his adjacent drivers. Also, tampering with air conditioning system, radio volume and distorting the dashboard [10], [11], [18], [20] not only compromise the traffic safety but leave a driver in a traumatic state, a state that can contribute to impulsive behavior.

Beyond the concerns mentioned above, a malicious adversary may compromise the road users' privacy through tracking attack. Such an attack can deteriorate if coupled with malware installation. Compromising in-vehicle network, impose monetary losses for road users and transportation operators – also, recuperating from an attack takes time. In-vehicle network vulnerabilities can cause users and government distrust. Moreover, individuals and companies might file a lawsuit against car manufacturer and authorities. Also, attacks can negatively affect the transportation efficiency by increasing the influenced vehicles' energy consumption [32]. Since there are not many studies to scrutinize attacks' consequences on road users and authorities, security attack impacts on transportation network shall be investigated more in-depth through stated and revealed methods (e.g., [33]–[38]).

The likelihood-impact matrix (Figure 3) represents the in-vehicle network security vulnerability impacts into a 3×3 matrix for risk prioritization purposes. The matrix links each of the adverse impacts to the three impact and likelihood clusters (low-, medium- and high-risk) to help decision-makers prioritize risks that are associated with hacked vehicles. Risk assessment result indicates that (i) safety (e.g., injury and fatality crash), operational (i.e., monetary loss), security (compromising driver's privacy based on the context), and behavioral (i.e., driver's distraction) impacts are associated with the high-risk cluster, and (ii) security (i.e., vehicles become a foothold in ITS), legal exposure, reliability (i.e., road users distrust the system), efficiency (i.e., passenger discomfort) and operational (i.e., congestion) impacts are associated with the medium-risk cluster. More research can supplement the risk analysis studies and help policymakers provide flexible solutions to mitigate the risk of cyber/physical attack regarding an in-vehicle network.

|  | LIKELIHOOD | | |
|---|---|---|---|
| IMPACT | Low | Medium | High |
| High | **Security Impact:** vehicles becomes a malicious node in the ITS network | **Safety Impact:** crashes with severity level of Injuries and fatalities<br>**Behavioral Impact:** traumatic experience that can contribute to an impulsive driver's behavior; Drivers' distraction | **Safety Impact:** property damage crashes<br>**Security Impact:** to compromise vehicles' availability; To compromise driver's privacy; to track road users' trajectories; Eavesdrop passengers' communications |
| Medium | **Legal Exposure:** due to losses, individuals and companies might file a lawsuit against manufacturer and authorities | **Operational Impact:** taking control of vehicle and compromise vehicle's critical components (e.g. engine, brake, steering wheel); Disabled and crippled services | **Operational Impact:** monetary loss of road users and system operators |
| Low |  | **Efficiency Impact:** increase in energy consumption | **Efficiency Impact:** cause discomfort to passengers<br>**Reliability Impact:** attacks on in-vehicle network can cause road users to distrust the intelligent vehicles, as well as the VANET future implementation |

Fig. 3. The Impact-Likelihood Matrix maps the adverse impact of in-vehicle network hacking into high- (colored in red), medium- (colored in yellow), and low-risk clusters (colored in green)

## VI. COUNTERMEASURES

This study objective is to raise awareness among transportation operators and traffic engineers to employ cybersecurity best practices to ensure:
(i) Security by design from inception,
(ii) Out-of-the-box solutions by examining the system as a whole not a collection of isolated components,
(iii) Redundancy solutions for critical cyber-physical systems,
(iv) Road users and operators familiarity with possible attacks and adequate training for them to react promptly to the attack, and
(v) A partnership among diverse resources (e.g., exercising white hackers), to bring innovative ideas and to portray the vulnerability of the system before the large-scale implementation.

The presence of any vulnerabilities triggers a race against the clock among adversaries and security experts. Thus, the in-vehicle network needs to be continuously assessed for potential vulnerabilities. Vulnerability scanning and penetration testing are among the primary solutions to assess the security and resiliency of the system against cyber threats [39]. Penetration testing is possible through sandboxing event and by inviting white hackers to identify system vulnerabilities in a controlled scenario. Besides, cyber hygiene; deployment of fault detection and isolation techniques (e.g., application of intrusion detection systems); and full confidentiality, integrity, availability, and authenticity threat analysis [39] are possible countermeasures that ensure security and resiliency of the in-vehicle network.

As stated earlier, CAN is the de facto communication protocol that connects ECUs within the in-vehicle network. ECUs are minicomputers, which have a specific task of automotive related duty. ECUs' tasks associated with, but not limited to, engine control, brake control, timing module. All the critical safety tasks are performed by ECUs that is, plausible to conclude that passenger's safety is dependent on the security and proper functionality of ECUs as stated in [14] "an attacker connected to a car's internal network can circumvent all computer control systems."

The deficiencies in automobile cyber-security stem from the fact that CAN protocol has a broadcast nature and does not support message confidentiality, authenticity or integrity. This provides the incentive for researchers to investigate the shortcomings of CAN protocol regarding security, and introduce reliable measures for automobiles. These contributions fall into three main categories, (i) protocols, (ii) intrusion detection systems (IDSs) and (iii) preventive methods, explained in detail below.

Different protocols have been proposed as an alternative to the conventional CAN protocol. Cryptographic solutions have been immensely investigated to secure CAN protocol as far as the authenticity and integrity measures are concerned. Message authentication codes (MACs) and digital signatures are some of the few approaches [40] to be named here. CAN+ [41] for instance, tries to overcome the problems of cryptographic methods by using out of band channels for transmitting the cryptographic bits. However, the main drawback of this scheme and similar solutions is the backward compatibility issue. Nilsson et al. [42] attempted to overcome the backward compatibility problem by authenticating multiple messages, which is less than optimal due to the use of multiple messages, which gives space to hackers for the intrusion. While Cryptographic methods provide secure solutions for CAN protocol, they may not be sufficient for providing a robust security mechanism. The reason is mainly that standard CAN frame only accommodates up to eight bytes. That is, CAN Cryptographic solutions might be vulnerable to the brute force attacks.

The second line of defense that offers solutions to the CAN security vulnerability problem includes IDSs. Unlike protocol security measures, IDSs can be mounted into the current CAN bus as a software patch that is completely aligned with the backward compatibility requirement. The state-of-the-art IDS mechanisms rely on a technique named fingerprinting. Fingerprinting generates templates for the network legitimate devices and uses the templates for identification purposes [43] – these feature templates are called fingerprints. In brief, any message that is to be transmitted into the CAN bus needs to match to the fingerprint of its source device or ECU. Otherwise, it will fail the identification step and will not be transmitted. Since fingerprints are usually extracted from the analog physical layer signal, this type of identification is called, physical layer identification (PLI) systems [43]. Examples of PLI based IDSs can be found in the literature that uses different features such as the line voltage [44]–[46] or clock skews [47] for extracting fingerprint templates.

Last, not least, it is evident that the automobile's security can be compromised through physical access [13] or a wireless connection [14]. The former is easily avoidable and brings in minor technical concerns, whereas the wireless connections established from the uncertified users are a matter of concern. Securing the wireless links is a problem that has been under-evaluated. So far, the telematics port or Bluetooth connection have been the primary source of security vulnerabilities. That is, footholds for adversaries to gain access to an automobile to inject malicious messages. Spaur et al. proposed a new telematics unit with certain access restrictions at [48] aiming to control the flow of messages into the vehicle through the cellular network. Based on their approach, this telematics ports is equipped with a security controller, which is built based on digital certificates. The controller grants access to certified users only. The security of telematics ports, even though essential, has been neglected in the current literature. The telematics ports have always been the most common and accessible footholds for malicious adversaries. To preclude intrusions right at the starting point, in-built security solutions –rather than reactive response– should be implemented.

The contributions to the field of automobile cybersecurity are well summarized in the three categories above. However, a robust framework offering a secure end-to-end physical platform is noticeably missing in state-of-the-art research. Providing awareness in public, as well as among professionals is of major importance, to bring the necessary attention to the transportation cybersecurity problem. The more significant problem, however, appears when security and privacy are taken into account at the same time. These two critical aspects of transportation systems are most of the time aligned in same directions and occasionally in opposite directions where findings an optimal solution for this tradeoff is very complicated.

## VII. CONCLUSION

ITS projected to augment safety, comfort, transportation efficiency, and to overcome the environmental impacts of transportation, but in the presence of security footholds all can dim. In the presence of any weaknesses, attackers can exploit ITS availability, authentication, integrity, and privacy security goals. Moreover, many questions need to be resolved before the vehicular ad hoc network large-scale implementation. For instance: (i) what would be the impact of known/unknown cyber/physical attacks on transportation? (ii) How should road users respond to the vehicle hacking incidents? (iii) Do road users need to be trained for cyber intrusion events? Moreover, (iv) Whom should road users reach for fixing their cars to recover from a security attack?

Security is a core technical challenge of the ITS that if not appropriately maintained, will drastically affect road users and system operators. Lessons learned from this study depict that road users' safety, privacy and security can be compromised due to the vehicle security vulnerabilities. Adversaries not only can take control of vehicles but can compromise the traffic safety and leave a driver in a traumatic state. Eventually, adversaries will supersede drivers to remotely-drive vehicles shortly.

This study offers an early assessment of the risks to the in-vehicle network. We employed the NIST risk assessment approach for risk determination. The risk assessment methodology conducted in identifying threats to the system, system vulnerabilities, adverse impacts in case an adversary exploits those vulnerabilities, and the likelihood that impacts will occur. To the best of the authors' knowledge, this study is the first attempt to systematically assess the risks of a compromised in-vehicle network in the transportation domain. The current literature is mainly devoted to detecting security threats and to suggesting protection measures against the identified and future security attacks. While the knowledge is lagging behind the state-of-the-art, the need for an impact-oriented analytic assessment to identify system-level risks is of vital importance. That is, this study aims to contribute to the body of knowledge of ITS security and resiliency by employing a qualitative risk-based approach for measuring risks and communicating the results for policy analysis. We summarize the main findings of the study as follows:

- An adversary can compromise the in-vehicle network due to the CAN vulnerabilities such as weak access control, no integrity and no authenticity. That is, a malicious adversary with high-level of expertise, well resourced, and capable of executing effective, continuous, and synchronized cyber/physical attacks, can exploit vehicle security vulnerabilities. As far as the connected/automated vehicle technology is concerned, these security vulnerabilities can accrue with VANET security issues. The security attack might extend and contaminate other vehicles that are on the same network as the compromised vehicle. Breach propagation can disseminate over the ITS wide-range network like a virus goes from one computer to another.
- The risk determination of an adversary compromising the in-vehicle network can be categorized into high-, medium-, and low-level risk clusters:
    - *High-risk cluster*: safety (e.g., injury and fatality crash), operational (i.e., monetary loss), security (compromising driver's privacy based on the context), and behavioral (i.e., driver's distraction) impacts.
    - *Medium-risk cluster*: security (i.e., vehicles become a foothold in ITS), legal exposure, reliability (i.e., road users distrust the system), efficiency (i.e., passenger discomfort) and operational (i.e., congestion) impacts.
- A rigorous solution offering a secure end-to-end physical platform is noticeably missing in state-of-the-art research. Providing awareness in public, as well as among professionals is of vital importance, to bring the necessary attention to the transportation cybersecurity problem. The more significant problem, however, appears when security and privacy are considered at the same time. Security and privacy aspects of the transportation systems are most of the time aligned in same directions and occasionally in the opposite directions where offering an optimal solution to this tradeoff is a complex task.

This study indicates a high level of uncertainty in the impact of a security breach on the in-vehicle network, especially on road users and the transportation network. There is a gap in literature to assess and evaluate the impact of cyber/physical attacks on the road users and the system operators. The result of this study endorses the necessity of impact assessment on the in-vehicle network security breach. Impact assessment results will be of great benefit to the risk management studies for the provision of realistic solutions against security attacks. Incorporating the outcomes of the impact assessment and the risk management provides a realistic analysis of the risk evaluation. Also, the impact assessment will help policymakers and system operators to prioritise strategies to mitigate security risk.